\documentclass[letterpaper, 10 pt, conference]{ieeeconf}  
\IEEEoverridecommandlockouts
\usepackage[utf8]{inputenc}
\usepackage{graphicx}
\usepackage{array}
\usepackage{epsfig}
\usepackage{amsfonts,amssymb,amsmath}
\usepackage{gensymb}
\usepackage{subcaption}
\usepackage{color}
\usepackage[ruled,vlined,titlenotnumbered]{algorithm2e}
\usepackage{algpseudocode}
\usepackage[colorlinks=true,citecolor=black,linkcolor=black,urlcolor=black]{hyperref}
\usepackage{multicol}
\usepackage{enumerate}

\graphicspath{{figures/}}
\DeclareGraphicsExtensions{.pdf,.jpeg,.png,.eps}

\makeatletter

\newcommand{\squeeze}[1]{{\vspace{-0.1cm}}}
\newcommand{\remove}[1]{}

\usepackage[textwidth=1.5cm]{todonotes}
{}
{}
{}
{}
{}
\newcommand{\reachset}{\mathcal{R}}
\newcommand{\targetset}{\mathcal{L}}
\newcommand{\traj}{\zeta} % trajectory
\newcommand{\pcfset}{\mathbb{U}_p} %planner control function set
\newcommand{\tcset}{\mathcal{U}_s} %tracker control set
\newcommand{\tcfset}{\mathbb{U}_s} %tracker control funciton set
\newcommand{\dset}{\mathcal{D}}

\newcommand{\pset}{\mathcal{P}} %planner set set
\newcommand{\tset}{\mathcal{S}} %tracker set

\newcommand{\tvar}{t}
\newcommand{\thor}{t_f} % Time horizon
\newcommand{\tstate}{s} % Tracker state
\newcommand{\pstate}{p} % Planner state
\newcommand{\rstate}{r} % Relative state
\newcommand{\ttraj}{\xi_{\tdyn}} % Tracker trajectory
\newcommand{\tctrl}{u_s} % Tracker control
\newcommand{\dstb}{d} % Disturbance
\newcommand{\pctrl}{u_p} % Planner control
\newcommand{\tdyn}{f} % Tracker dynamics
\newcommand{\pdyn}{h} % Planner Dynamics
\newcommand{\rdyn}{g} % Relative dynamics

\newcommand{\ptmat}{Q} % Matrix for transforming planner state to the same length as tracker state

\newcommand{\TEB}{\mathrm{TEB}} % tracking error bound
\newcommand{\To}{\longrightarrow}

\def\Vec#1{\!\!\hbox{$#1$\kern-0.38em\lower0.85em\hbox{$\vec{}\,$}}\,}%
\newcommand{\bbm}{\begin{bmatrix}}
\newcommand{\ebm}{\end{bmatrix}}
\newcommand{\bpm}{\begin{pmatrix}}
\newcommand{\epm}{\end{pmatrix}}

\newcommand{\mc}[1]{\mathcal{#1}}

\newtheorem{remark}{Remark}

\newtheorem{theorem}{Theorem}

\newtheorem{proposition}[theorem]{Proposition}

\usepackage{balance}

\usepackage[%
    style=numeric-comp,
    sorting=none,
    backend=biber,
    sortcites=true,
    doi=false,
    firstinits=true,
    hyperref,
    isbn=false,
    eprint=false,
    maxcitenames=3, %How many people before *choosing* to contract with [et al.]
    minbibnames=3, %How many authors *at least*  before *inserting* the [et al.]
    maxbibnames=4, %How many authors *at most*  before *inserting* the [et al.]
    block=none]
    {biblatex}
    
    \renewbibmacro{in:}{}
    \AtEveryBibitem{%
  	\clearlist{language}%
  	\clearfield{pages}%
	}
\bibliography{references} % Jaime

\title{Planning, Fast and Slow: \\ A Framework for Adaptive 
Real-Time Safe Trajectory Planning}
\author{
\authorblockN{David Fridovich-Keil*, Sylvia L. Herbert*,  Jaime F. Fisac*, Sampada Deglurkar, and Claire J. Tomlin}
\thanks{*These authors contributed equally. This research is supported by NSF under the CPS Frontiers VehiCal project (1545126), by the UC-Philippine-California Advanced Research Institute under project IIID-2016-005, and by the ONR MURI Embedded Humans (N00014-16-1-2206). S. Herbert and D. Fridovich-Keil are funded by the NSF GRFP. S. Herbert is also funded by the UC Berkeley Chancellor’s Fellowship. \{dfk, sylvia.herbert, jfisac, sampada\_deglurkar, tomlin\}@berkeley.edu} %{\tt\small\{dfk, sylvia.herbert, jfisac, sampdata\_deglurkar, tomlin\}@berkeley.edu}}%, sampada\_deglurkar@berkeley.edu}}
}

\begin{document}

\maketitle \thispagestyle{empty} \pagestyle{empty}

% ===================== ABSTRACT =====================
\begin{abstract}
Motion planning is an extremely well-studied problem in the robotics community, yet existing work largely falls into one of two categories: computationally efficient but with few if any safety guarantees, or
% inefficient but with stronger guarantees
able to give stronger guarantees but at high computational cost. This work builds on a recent development called FaSTrack in which a slow offline computation provides a modular safety guarantee for a faster online planner. We introduce the notion of ``meta-planning'' in which a refined offline computation enables safe switching between different online planners. This provides autonomous systems with the ability to adapt motion plans to \textit{a priori} unknown environments in real-time as sensor measurements detect new obstacles, and the flexibility to maneuver differently in the presence of obstacles than they would in free space, all while maintaining a strict safety guarantee. We demonstrate the meta-planning algorithm both in simulation and in hardware using a small Crazyflie 2.0 quadrotor.
\end{abstract}

% ==================== INTRODUCTION ==================
\section{Introduction}
\label{sec:introduction}

\begin{figure}[t]
\centering
\includegraphics[width=0.85\columnwidth]{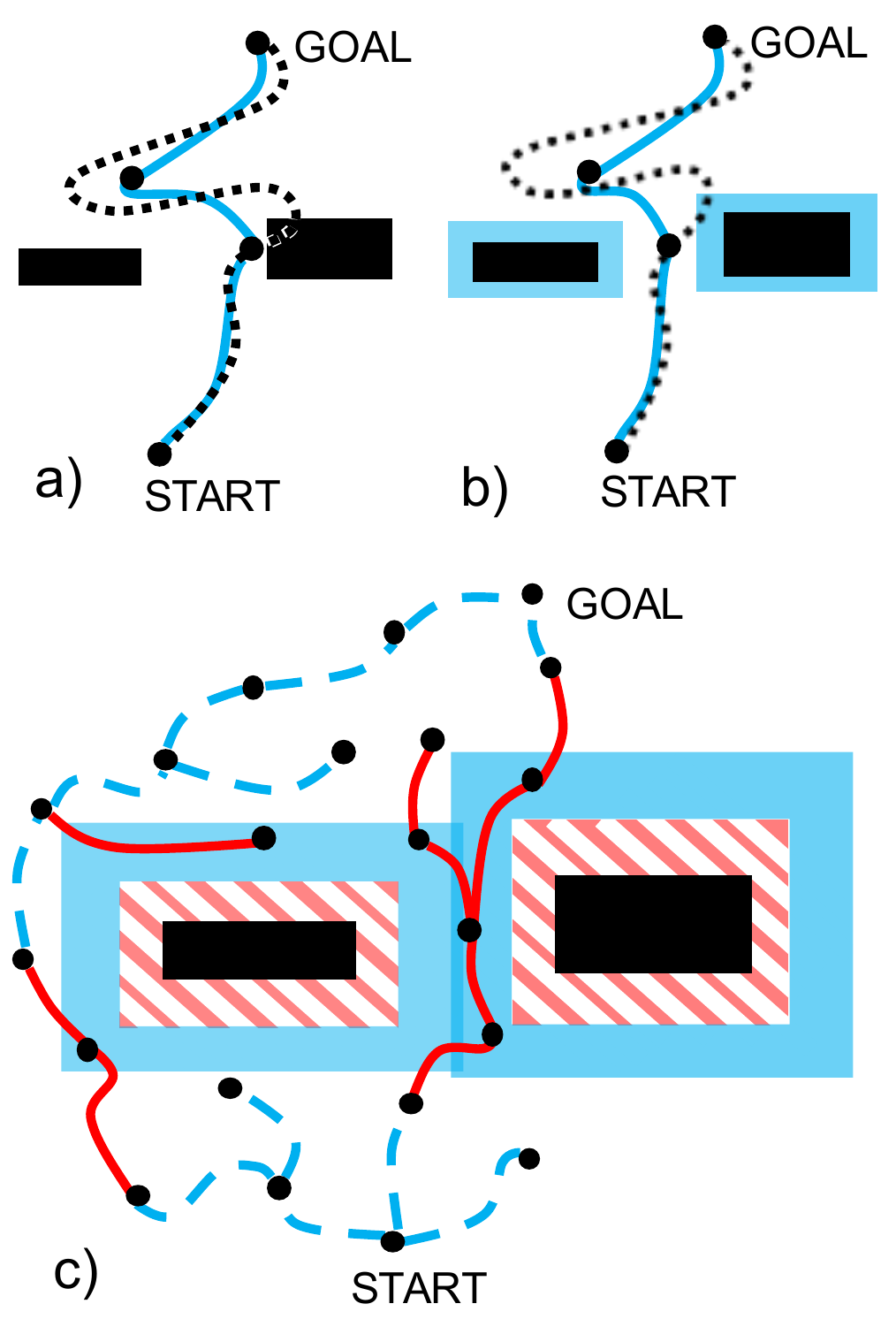}
\caption{
(a)~A dynamical system (black, dotted) may not be able to track the output of a geometric planner (blue, solid), resulting in collision with an obstacle.
(b)~Often planners account for tracking error by heuristically augmenting obstacles; however, the system may still deviate from the planned path by more than this margin.
(c)~Schematic of meta-planner operation using fast (blue, dashed) and slow (red, solid) planning models with correspondingly large (blue, solid) and small (red, hatched) TEB-augmented obstacles.}
\label{fig:Tree}
\vspace{-.5cm}
\end{figure}

The navigation of autonomous dynamical systems through cluttered environments is a hard problem, particularly when there is a need for both speed and safety. Often, elements of the environment (such as obstacle locations) are also unknown \textit{a priori}, further complicating the problem. Many popular methods exist for planning trajectories in such scenarios, but a key challenge lies in accounting for dynamic feasibility in real time while providing a safety guarantee. Some of the most common approaches in this space are sampling-based planners such as rapidly-exploring random trees (RRTs) \cite{lavalle1998rapidly}. Typically, these planners fall into one of two broad categories: \textit{geometric} planners only attempt to find a \textit{path} the system can take from its current position to the goal, while \textit{kinodynamic} planners find a dynamically feasible \textit{trajectory}, i.e. a path with associated time stamps that adheres to some known system dynamics.
%Though both types of planning are widely used, in this work we focus on geometric planners, since they are typically more computationally tractable; however, virtually all of our results hold for kinodynamic planners as well. 

Since the output of a geometric planner is not usually dynamically feasible, a common practice is to apply a feedback controller, e.g. a linear quadratic regulator (LQR), to attempt to track a geometric plan. Since the controller will not follow the plan perfectly, geometric plans are usually generated by assuming an ad hoc safety margin.
This idea is illustrated in Fig.~\ref{fig:Tree}(a-b).
%, which shows how introducing a safety margin around obstacles prevents a geometric planner from generating a trajectory resulting in collision for the real system.

% \begin{figure}[t]
% \centering
% \includegraphics[width=0.48\textwidth]{figures/Overshoot.pdf}
% \caption{A dynamical system (red, dashed) may not be able to track the output of a geometric planner (blue, solid). In (a), this results in collision with an obstacle, but in (b) the planner accounts for tracking error by augmenting the obstacles, resulting in a collision-free system trajectory.}
% \label{fig:Overshoot}
% \end{figure}

In practice, this safety margin is almost always a \textit{conservative heuristic} chosen by the operator. However, the recently-developed Fast and Safe Tracking (FaSTrack) framework \cite{Herbert2017} provides a rigorous way to precompute a safety margin offline, given a model of the true system dynamics and a (possibly lower-dimensional) model of the online planner's dynamics. In the FaSTrack framework, a guaranteed maximum possible tracking error is computed between the tracking system model and the planning model. This tracking error bound (TEB) can also accommodate deviations due to external disturbances such as wind and time delays. The TEB is used to expand obstacles by a margin that \emph{guarantees} safety. The offline precomputation also provides a computationally efficient safety controller that maps the relative state between the tracking system and the planned trajectory at any given time to the most effective control action for the tracking system to remain within the TEB. Hence, the online algorithm involves real-time planning using a fast, potentially low-dimensional planning model, and quickly computable robust optimal tracking of the planned trajectory using a higher-dimensional tracking model.

While FaSTrack makes no significant assumptions about the specific type of low-dimensional planner, in this work we focus our attention on geometric planners operating in the robot's configuration space.
We observe that the resulting geometric paths can be interpreted as kinematic trajectories with a fixed maximum speed in each dimension.
We emphasize that the restriction to geometric planners is pedagogical; like FaSTrack, our proposed meta-planning approach is more general and extends to more complex planning models.

One key drawback of FaSTrack is that the TEB can be overly conservative if the system is tracking a particularly difficult-to-track planning model. In this paper we propose an extra layer to the core framework that allows combining multiple planning models with different maximum speeds, and hence different TEBs. We call this process \textit{meta-planning}, and it  effectively generates a tree of trajectories that switch between ``faster'' and ``slower'' planning models, as illustrated in Fig. \ref{fig:Tree}(c). Faster planning models are able to navigate through the environment quickly, but their larger TEBs prevent them from threading narrow passages between obstacles. Slower planning models take more time to traverse the environment, but the correspondingly smaller TEBs allow them to maneuver more precisely near obstacles. By adaptively selecting the planning model in real time, our framework can trade off between speed of navigation and size of the TEB. Crucially, our meta-planning scheme can quickly and \emph{safely} adapt to the presence of obstacles detected at motion time.

% TODO! Make sure we do have a constructive proof of safety.
The main contributions of this paper are the aforementioned real-time meta-planning algorithm for Fast and Safe Tracking, a constructive proof of safety, and a demonstration of the full algorithm both in simulation and hardware using a small quadrotor vehicle.

%in the form of a meta-planner.  This meta-planning layer considers a suite of available planners, and adaptively selects the planner best suited for a given scenario in real time based on order of preference.  Note that because the FaSTrack algorithm is agnostic to type of planner \footnote{as long as the planner can be represented by a model, see section \ref{sec:background}}, this suite may contain a variety of types of planners.  For example, in a wide-open area one might prefer to use a planner that is fast but difficult to track, whereas in a cluttered space a slower but more maneuverable planner may be best for the job.

% ==================== INTRODUCTION ==================
\section{Related Work}
\label{sec:related_work}

Robust motion planning and trajectory optimization have been active areas of research in recent years. However, navigation that is both robust and fast is still a challenge.  Sampling-based motion planners can be computationally efficient, but attempts to make them robust are generally heuristic. Other techniques for online dynamic navigation include model predictive control (MPC), which is extremely useful, particularly for linear systems. MPC is harder to use in real time for nonlinear systems due to the computational costs of solving for dynamic trajectories, though work to speed up computation is ongoing \cite{Diehl2009, Neunert2016}. Robustness can be achieved in linear systems \cite{Richards2006, DiCairano2016}, and there is work on making MPC for nonlinear systems robust by using algorithms based on minimax formulations and tube MPCs that bound output trajectories with a tube around a nominal path (see \cite{Hoy2015} for references).

There are other techniques for robust navigation that take advantage of precomputation. Safety funnels can be constructed around motion primitives that can then be pieced together in real time~\cite{Majumdar2017}.  Given a precomputed nominal dynamically feasible trajectory, contraction mapping can be used to make this nominal trajectory more robust to external disturbances in real time~\cite{Singh2017}.
Finally, Hamilton-Jacobi (HJ) reachability analysis has been used for \emph{offline} robust trajectory planning in fully known environments, providing guaranteed tracking error bounds under external disturbances~\cite{Bansal2017}.

The meta-planning aspect of this paper was inspired by behavioral economist Daniel Kahneman's Nobel Prize winning work on ``fast'' (intuitive) and ``slow'' (deliberative) modes of cognitive function in the brain \cite{kahneman2011thinking}. Thinking with the ``fast system'' is efficient, but more error-prone. Thinking with the ``slow system'' is less error-prone, but slower. The brain adaptively chooses which mode to be in to operate efficiently while minimizing error in scenarios where error can be disastrous. This act of deciding how much cognitive effort to expend for a given task is called metareasoning \cite{russell1991principles}, and can be useful for robotics. It may be desirable for a robot to plan and move swiftly whenever possible, but to operate more carefully when approaching a challenging region in the environment. Research in psychology has suggested that selecting between a limited number of discrete cognitive modes is computationally advantageous \cite{milli2017does}, which inspires the use of discrete set of faster and slower planning models in our meta-planning algorithm.
%In addition to being robust, efficiency in navigation is important.  Researchers are working on methods for hierarchical motion planning or continuous refinement of motion planning \cite{Ratliff2009} \SHnote{struggling to find examples of people selecting among multiple planners in real time}.
Our algorithm is able to trade off planner velocity and tracking conservativeness in a modular way while providing a strong theoretical safety guarantee. % while maintaining safety,
%FaSTrack uses simple and potentially dynamically infeasible trajectories that are planned online using \textit{any} planner that can be represented by a dynamic model.
%generating trajectories that can be tracked in real time via FaSTrack, in a dynamically feasible way that is robust to disturbances and has theoretical safety guarantees.
%In this paper we focus on the use of FaSTrack for tracking geometric planners, and contribute the novel meta-planning algorithm for trading off between fast and slow planners in real time while continuing to maintain safety.

% ==================== BACKGROUND ====================
\section{Background}
\label{sec:background}

The FaSTrack framework can be used to plan and track a trajectory online and in real time. The real-time planning is done using a set of kinematic or dynamic planning models, and the physical system is represented by a dynamic tracking model that will attempt to follow the current planning model. The environment can contain static \textit{a priori} unknown obstacles provided they can be observed by the system within a limited sensing range.\footnote{In order to provide safety guarantees, the minimum allowable sensing distance in any direction is the length of the TEB's projection onto that direction, added to the largest distance the current planning reference could move while a new meta-plan is generated.} In this section we will define the tracking and planning models and their relation to one another, and present a brief overview of FaSTrack.

\subsection{Tracking Model}
\label{subsec:tracking_model}

The tracking model should be a realistic representation of the real system dynamics, and in general may be nonlinear and high-dimensional. Let $\tstate$ represent the state variables of the tracking model. The evolution of the dynamics satisfies the ordinary differential equation (ODE): 
\begin{equation}
\begin{aligned}
\label{eq:tdyn}
\frac{d\tstate}{d\tvar} = \dot{\tstate} = \tdyn(\tstate, \tctrl, \dstb), \tvar \in [0, \thor] \\
\tstate \in \tset, \tctrl \in \tcset, \dstb \in \dset
\end{aligned}
\end{equation}
%We assume that the system dynamics $\tdyn : \tset\ \times\ \tcset \times \dset \rightarrow \tset$ are uniformly continuous, bounded, and Lipschitz continuous in $\tstate$ for fixed control $\tctrl$. %The control function $\tctrl(\cdot)$ and disturbance function $\dstb(\cdot)$ are drawn from the following sets:
%\begin{equation}
%\begin{aligned}
%\tctrl(\cdot) \in \tcfset(t) = \{\phi: [0, \thor] \rightarrow %\tcset: \phi(\cdot) \text{ is measurable}\}\\
%\dstb(\cdot) \in \dfset(t) = \{\phi: [0, \thor] \rightarrow %\dset: \phi(\cdot) \text{ is measurable}\}
%\end{aligned}
%\end{equation}
%where $\tcset, \dset$ are compact and $t\in[0, \thor]$ for some $T>0$. Under these assumptions there exists a unique trajectory solving (\ref{eq:tdyn}) for a given $\tctrl(\cdot) \in \tcset$ \cite{Coddington84}. 
The trajectories of (\ref{eq:tdyn}) will be denoted as $\ttraj(\tvar; \tstate_0, \tvar_0, \tctrl(\cdot), \dstb(\cdot))$, where $\tvar_0,\tvar \in [0, \thor]$ and $\tvar_0 \leq \tvar$. 
Under standard technical assumptions~\cite{Herbert2017}, these trajectories will satisfy the initial condition and the ODE (\ref{eq:tdyn}) almost everywhere.
% \begin{equation}
% \label{eq:fdyn_traj}
% \begin{aligned}
% \frac{d}{d\tvar}\ttraj(\tvar; \tstate, \tvar_0, \tctrl(\cdot)) &= \tdyn(\ttraj(\tvar; \tstate, \tvar_0, \tctrl(\cdot)), \tctrl(\tvar)) \\
% \ttraj(\tvar; \tstate, \tvar, \tctrl(\cdot)) &= \tstate
% \end{aligned}
% \end{equation}
For a running example we will consider a tracking model of a simple double-integrator with control $\tctrl$ and disturbances ${\dstb = [\dstb_v, \dstb_{a}]^T}$:
\begin{equation}
\small
\label{eq:tracking_dyn_example}
\begin{aligned}
\begin{array}{c}
\left[
\begin{array}{c}
\dot{\tstate}_x\\
\dot{\tstate}_{vx}\\
\end{array}
\right]
=
\left[
\begin{array}{c}
\tstate_{vx} - d_v\\
\tctrl -d_{a}\\
\end{array}
\right]
\end{array}
\end{aligned}
\end{equation}

\subsection{Planning Model}
\label{subsec:planning_model}
The planning model defines the class of trajectories generated by the motion planner. Let $\pstate$ represent the state variables of the planning model, with control $\pctrl$. The planning states $\pstate \in \pset$ are a subset of the tracking states $\tstate \in \tset$. FaSTrack is agnostic to the type of planner, as long it can be represented using a kinematic or dynamic model as follows:
\begin{equation}
\begin{aligned}
\label{eq:pdyn}
\frac{d\pstate}{d\tvar} = \dot{\pstate} = \pdyn(\pstate, \pctrl), \tvar \in [0, \thor], \pstate \in \pset, \ \underline{\pctrl} \leq \pctrl \leq \overline{\pctrl}
\end{aligned}
\end{equation}
This paper focuses on geometric planners. Although geometric planners may not directly use a dynamical model, the paths they generate can be described by a point with direct velocity control. For example, a 1D geometric planner could be described as a point moving with a direct velocity controller: $\dot{\pstate}_x = \pctrl$.
Note that the planning model does not need a disturbance input. Disturbances need only be considered in the tracking model and not the planning model, since the latter only exists in the abstract as a reference for the former.%, which is modular with respect to any planning method, including those that do not account for disturbances. 

\subsection{Relative Dynamics}
\label{subsec:relative_dynamics}
The FaSTrack framework relies on using the relative dynamics between the tracking and planning models. The relative system may be derived by lifting the planner's state from $\pset$ to $\tset$ and subtracting:
\begin{equation}
\label{eq:rdyn}
\begin{aligned}
\rstate = \tstate - \ptmat\pstate, \qquad \dot\rstate = \rdyn(\rstate, \tctrl, \pctrl, \dstb)
\end{aligned}
\end{equation}

\noindent $\ptmat$ is matrix that matches the common states of $\tstate$ and $\pstate$ by augmenting the state space of the planning model. The relative states $\rstate$ now represent the tracking states relative to the planning states. Using our tracking and planning model examples from above we can define the dynamics of a double-integrator tracking a 1D point mass as:%Similarly, $\tpmat$ projects the state space of the tracking model onto the planning model: $\pstate = \tpmat(\tstate-\rstate)$. This will be used to update the planning model in the online algorithm.

\begin{equation}
\small
\label{eq:rdyn_example}
\begin{aligned}
\begin{array}{c}
\left[
\begin{array}{c}
\dot{r}_x\\
\dot{s}_{vx}\\
\end{array}
\right]
=
\left[
\begin{array}{c}
\tstate_{vx} - d_v - \pctrl\\
\tctrl -d_a\\
\end{array}
\right]
\end{array}
\end{aligned}
\end{equation}

\subsection{The FaSTrack Framework}
\label{subsec:fastrack}
The FaSTrack framework, explained in detail in \cite{Herbert2017}, consists of both an offline precomputation algorithm and an online planning algorithm. Together, these allow a nonlinear dynamic system to navigate through an \textit{a priori} unknown environment with static obstacles, safely and in real time.

Offline, FaSTrack computes a tracking error bound (TEB) and a safety controller to stay inside this bound. The TEB is a safety margin that, when using the safety controller, guarantees robust tracking despite worst-case planner behavior and bounded disturbances.
%The relative state between the tracking and planning systems is computed at each time step and input to a safety controller, which outputs a control signal to the autonomous system. 
The safety controller operates on the \textit{relative} state between tracker and planner, and is computed offline via HJ reachability analysis \textit{in free space}. This is possible because the relative dynamics do not depend on the absolute state of the tracking system in the environment. Since the tracker will always remain inside the TEB, as long as the TEB never intersects any obstacles, the free space relative dynamics will always apply.
%\SHnote{Relate level set to TEB more precisely either here and/or the intro}
%To precompute the TEB and the corresponding safety controller, the relative dynamics are used in a HJ reachability computation to determine the tracking error value function $\valfunc(\rstate)$ over a discretized state space. The minimum level set of this function represents the smallest invariant set that the tracker can stay within relative to the planner, and its projection onto the position/configuration space is the TEB. The gradient of this function $\deriv(\rstate)$ determines the safety controller. 
%These functions depend only on the relative dynamics and not absolute states or dynamics of the systems at execution time. This allows FaSTrack to be used for real-time generation of trajectories \comment{in arbitrary environments.}

Online, both at the start and whenever a new obstacle is sensed, an off-the-shelf planning algorithm---equipped with the precomputed TEB for collision-checking---generates a new trajectory. The tracking system may then apply the precomputed safety controller to track this planned trajectory in real time.

% =================== META PLANNING ==================
\section{Meta-Planning}
\label{sec:meta_plannning}

\subsection{General Framework}
\label{subsec:metaplanning_general_framework}
In this work, we use the term \emph{planner} to denote the conjunction of a planning algorithm and an associated planning model that it uses to generate timed trajectories.%
This paper's main contribution to the FaSTrack framework is the introduction of a meta-planning algorithm to choose between a selection of planners $\{\pi_i\}_{i=1}^N$ with different maximum speeds and hence different TEBs at runtime. We first assume that planners are sorted in order of decreasing maximum speed and hence TEB size, and that the overall objective is to minimize the time to reach a specified goal point. This objective implies a preference for planners that can move faster, but also for planners that can safely navigate a more direct route even if they must do so at lower speed. 

The core of the meta-planner is a random tree $\mc{T}$ inspired by RRT-style sampling-based planners \cite{lavalle1998rapidly}, as shown in Fig.~\ref{fig:Tree}.\footnote{The choice of a tree topology is for convenience; any directed graph %---e.g. a probabilistic roadmap (PRM) \cite{Kavraki1996}---starting at
including the robot's current state would suffice.} The obstacles are shown in black, and are augmented by the TEBs for two different planners. As in RRT, waypoints in $\pset$ are sampled randomly from the environment and (potentially) connected with their nearest neighbor in $\mc{T}$. If the fast planner%
\footnote{Note that by a \emph{faster} planner we mean one with a higher associated maximum velocity, rather than smaller computation time.}
(with the large blue TEB) finds a collision-free trajectory, the connection is established (dashed blue lines). Otherwise, the slow planner (with smaller red striped TEB and solid red lines) attempts to connect to the nearest neighbor. Upon success, the waypoint is inserted into $\mc{T}$, along with the trajectory generated by the planner to reach that waypoint from the nearest neighbor, and the associated safety controller to remain inside the TEB. If a waypoint is successfully inserted near the goal, a similar process ensues to attempt to find a trajectory between it and the goal point. 

Once a valid ``meta-plan'' is found from start to goal, the meta-planner continues building $\mc{T}$ until a user-specified maximum runtime has elapsed, always retaining the best (shortest time) sequence of waypoints to the goal. Similar to Informed RRT* \cite{gammell2014informed}, the meta-planner immediately rejects samples
%which the fast planner could not reach
%(in straight lines from/to start and goal) 
%in less than the duration of the best available trajectory.
which could not possibly improve upon the best available trajectory.\footnote{In Informed RRT*, planner velocities lie on the sphere leading to an elliptical geometry for valid samples. Since we assume a maximum speed \textit{in each dimension}, valid samples lie in a distorted rhombicuboctahedron.}

The key to meta-planning lies in ensuring safe switching between planners. This guarantee requires an offline computation to determine a safety margin for switching into successively slower planners (with smaller TEBs), as well as a safe switching control law. Online, we must be sure to plan with the appropriate safety margin at each step, and to ``backtrack'' if we detect the need for a switch to a slower planner. We will next explore the offline and online steps in detail.%Offline computation is discussed in greater detail in Sec. \ref{subsec:metaplanning_offline}, while the online component is presented in Sec. \ref{subsec:metaplanning_online}.

%This tree then informs the FaSTrack algorithm on which planner to use between two waypoints, as well as when to switch between two planners. An analysis of how to ensure safe switching is in section \ref{sec:metaplanning_offline}. The FaSTrack algorithm is thus augmented to meta-FaSTrack, as shown in Fig.~\ref{fig:meta-FaSTrack_Framework}.

\subsection{Offline Reachability Analysis}
\label{subsec:metaplanning_offline}
\begin{figure}
\centering
\includegraphics[width=0.27\textwidth]{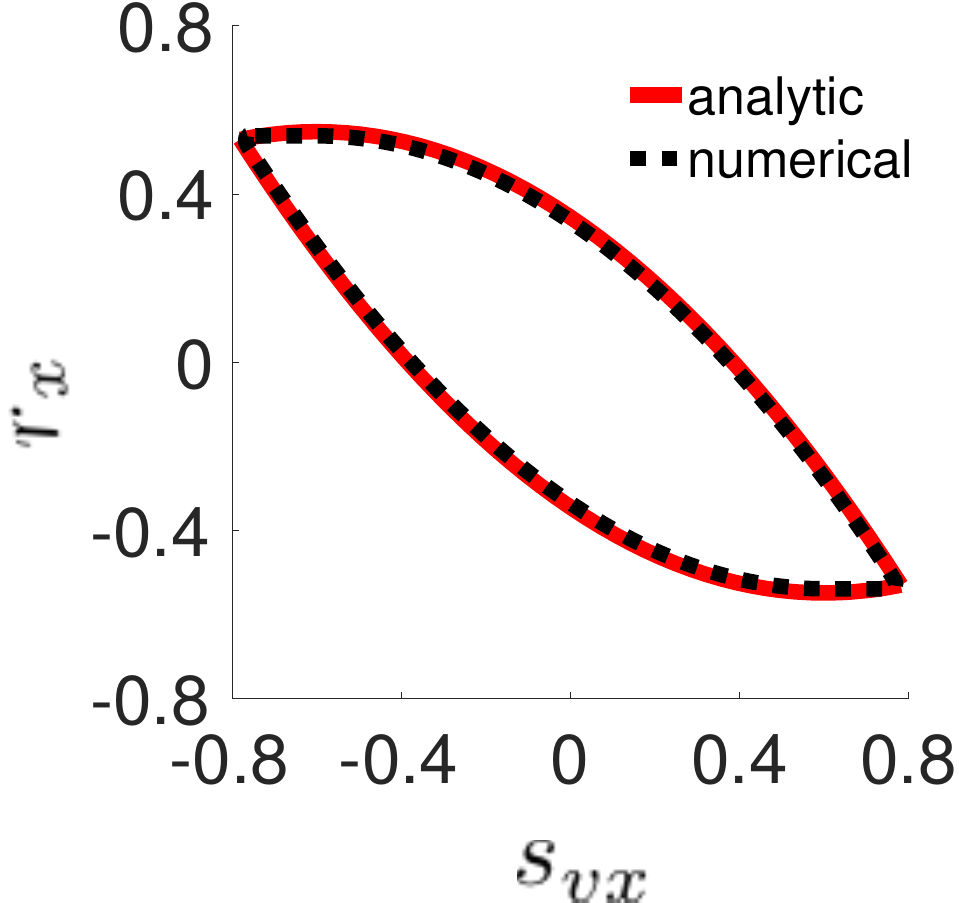}
	\caption{Invariant set that the double-integrator can remain in despite worst-case disturbance and planning control for the both numerical solution (dotted) and analytic solution (solid).}
	\label{fig:Analytic_vs_Numerical_TEB}
\end{figure}

There are two major components to the offline precomputation for the meta-planner. The first step is to compute the TEB and safety control look-up tables for each planner. This is done following the standard FaSTrack precomputation algorithm \cite{Herbert2017}. 
Fig.~\ref{fig:Analytic_vs_Numerical_TEB} shows the set of relative states in the $x$-subsystem that the tracker can remain within despite worst case planner behavior and external disturbance. The projection of this controlled invariant set onto the position axis comprises the $x$-TEB.
For the double-integrator dynamics in \eqref{eq:rdyn_example}, an analytic solution can also be found by applying the equations of constant-acceleration motion
under the worst-case disturbance and the best associated control effort.
% as the intersection of two parabolas defined by (\ref{eq:parabolas}), where the overbars represent the maximum value of the parameters.
The analytic controlled invariant set,
consisting of two parabolic curves,
is superimposed in Fig.~\ref{fig:Analytic_vs_Numerical_TEB}.
Such analytic solutions do not exist in general.

% \begin{equation}
% \label{eq:parabolas}
% \begin{aligned}
% c &= \pm \frac{1}{2}(v_x \pm \bar{\pctrl} \pm \bar{\dstb_v})^2 \pm (\bar{\pctrl} + \bar{\dstb_v})/(\bar{\tctrl} - \bar{\dstb_{a}})\\
% \end{aligned}
% \end{equation}

The second major component of the offline precomputation is to find the corresponding tracking bound and optimal controller for transitioning between planners. For the dynamics in \eqref{eq:rdyn_example}, switching from a planner with a small TEB to one with a large TEB is safe by construction, because the large TEB \emph{contains} the small TEB. Switching from a large TEB to a small one is more complicated.

\begin{figure}
\centering
\includegraphics[width=0.2\textwidth]{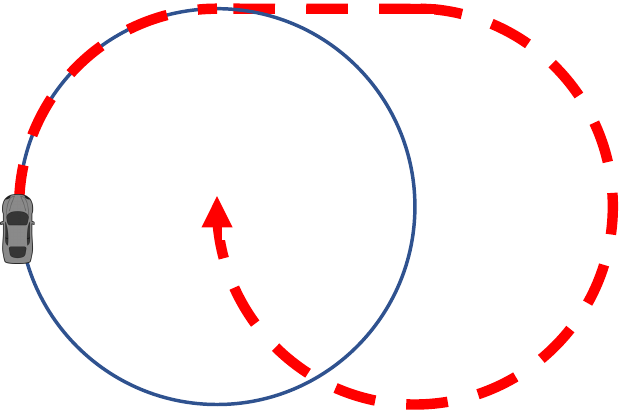}
	\caption{Example of a Dubins car that must leave its tight orbit in order to eventually move closer to the origin. This example illustrates why the switching safety bound may generally be larger than the tracking error bound.}
	\label{fig:Dubins_Example}
\end{figure}

To transition from a large TEB to a small TEB we must ensure that the relative state between the autonomous system and the planned path is within the small TEB by the time of the planner switch.
FaSTrack provides the optimal control for staying within each bound individually, but does not provide the controller and bound required for reducing the tracking error prior to a switch. Perhaps surprisingly, in general the tracker may first need to exit the large TEB before converging to the small TEB. Fig.~\ref{fig:Dubins_Example} provides an intuitive example of this phenomenon. Here, a Dubins car moving at a fixed speed remains within radius $R$ of the origin by turning at its maximum steering angle. In order for the car to reduce its distance to the origin, it must first exit the original circle to reorient itself towards the origin. In general we must precompute the set of states that the system may visit when transitioning from a large TEB to a small TEB, and the optimal control to achieve this transition. To do this we use HJ reachability analysis.

HJ reachability analysis provides a rigorous mechanism for analyzing the goal satisfaction of a system, and can be used to determine the backward reachable tube (BRT). The BRT is the set of all allowable initial states of a system such that it can enter a set of goal states within a given time interval. HJ reachability analysis can also be used in the context of differential pursuit-evasion games \cite{Huang11, Chen17}. Here, as in FaSTrack \cite{Herbert2017}, we assume there is such a game between the tracking system and the planning system. In this game, the tracking system will try to ``capture" the planning system, while the planning system is attempting to avoid capture. In practice, the planner is not actively trying to avoid the tracker, but this assumption accounts for unexpected, worst-case planner behavior. We want to determine the set of states that the tracking system may visit when transitioning from the larger TEB to the smaller TEB.

Before constructing the differential game we must first determine its information structure, i.e. how and when each player makes decisions.
Since the relative dynamics between the tracker and planner are decoupled in their respective inputs, and we assume an additive disturbance, it is in fact irrelevant who ``plays first" at each time instant, and the value of the game is well defined under feedback strategies.
% We give an advantage to the planner and disturbance by forcing the tracker to play first at each time. %step.
% The planner and disturbance then choose their inputs based on the choice made by the tracker. We restrict the planner and disturbance to use nonanticipative strategies: at each iteration they know the current action of the tracker, but no future actions.

%We define a strategy for planning system as the mapping $\gamma_{\pstate} : \tcset \rightarrow \pcset$ that determines a control for the planning model based on the control of the tracking model. We restrict $\gamma$ to draw from only non-anticipative strategies $\gamma_{\pstate} \in \Gamma_\pstate(t)$, as defined in \cite{Mitchell05}. We similarly define the disturbance strategy $\gamma_{\dstb}: \tcset \rightarrow \dset$, $\gamma_{\dstb} \in \Gamma_\dstb(t)$.

For the system in the form of \eqref{eq:rdyn}, we would like to compute the BRT of time horizon $T$, denoted $\reachset(T)$. Intuitively, $\reachset(T)$ is the set of states from which there exists a control strategy to drive the system into a target set $\targetset$ within a duration of $T$ despite worst-case disturbances. Formally, the BRT is defined here as%\footnote{Similar definitions of BRTs and their relationships can be found in, for example, \cite{Mitchell07}.}

\begin{equation}
\label{eq:reachset}
\begin{aligned}
\reachset(T) = \{&\rstate : \exists \tctrl(\cdot)\in\tcfset, \forall \pctrl(\cdot)\in\pcfset, \forall \dstb(\cdot)\in\dset,\\
&\rstate(\cdot) \text{ satisfies \eqref{eq:rdyn}},\\ 
&\exists \tvar \in[\tvar_0-T,\tvar_0], \traj(\tvar; \rstate, \tvar_0, \tctrl(\cdot), \dstb(\cdot)) \in \targetset\}
\end{aligned}
\end{equation}
where $\tcfset,\pcfset,\dset$ denote the sets of feedback strategies for the tracker, planner and disturbance.

\begin{figure}
    \centering
    \begin{subfigure}[t]{0.49\linewidth}
    \includegraphics[width=1\linewidth]{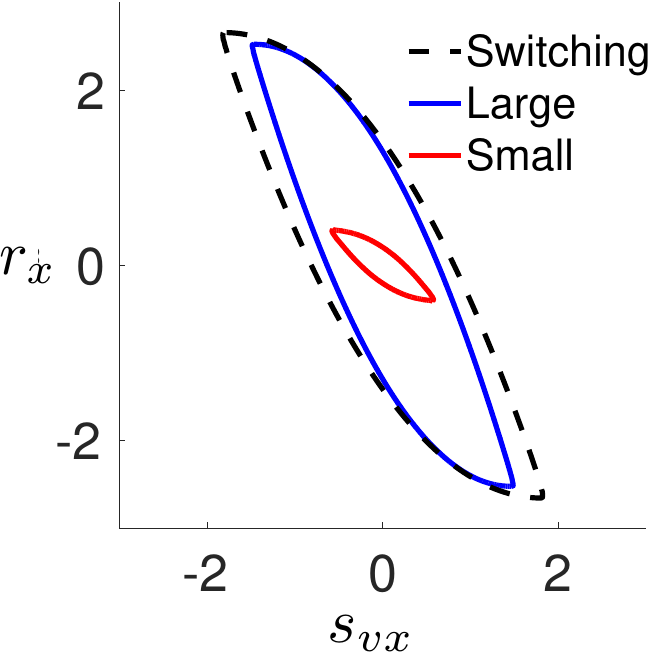}
    \end{subfigure}
    \begin{subfigure}[t]{0.49\linewidth}
        \includegraphics[width=1\linewidth]{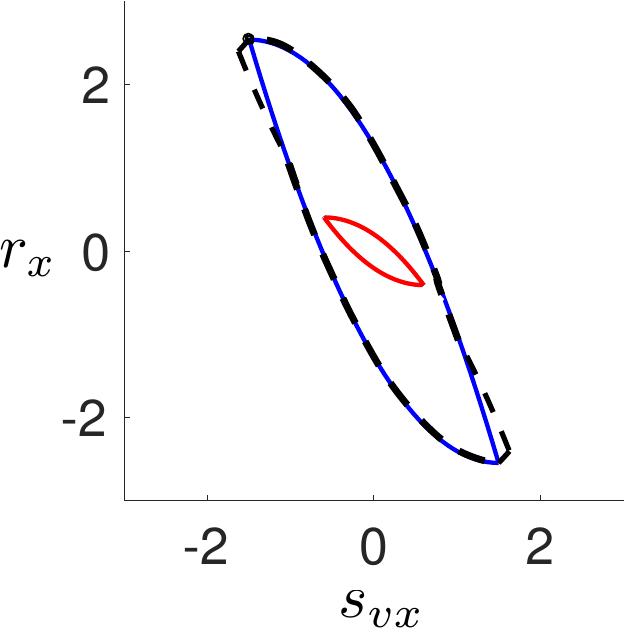}
    \end{subfigure}
    \caption{Visualizations of the $x$-subsystem's numerical (left) and analytic~(right) controlled invariant sets for two different planners. The numerical SSB is guaranteed to over-approximate the minimal %true
    SSB.}
    \label{fig:TEB_Switching_2D}
\end{figure}

Standard HJ formulations exist for computing the BRT in general \cite{Barron90, Mitchell05, Bokanowski10, Fisac15}, and more efficiently for decomposable systems \cite{Chen2016DecouplingJournal}. Here the target $\targetset$ is the set of states represented by the smaller tracking error bound.  Using the relative dynamics between the tracking model and the planning model associated with the smaller TEB, we evolve this set backwards in time.  We stop the computation when the tube contains the set of states associated with the larger TEB. This BRT represents the set of states from which the system can enter the small TEB, as well as the states that the trajectories may enter along the way. By projecting this set onto the position dimensions we obtain a %have the
\emph{switching safety bound} (SSB). We note that this is an \textit{over-approximation} of the minimal %true 
SSB because it includes
trajectories that do not originate inside
the larger TEB's controlled invariant set.
The SSB precomputation also generates the switching controller. Continuing our double-integrator example, Fig.~\ref{fig:TEB_Switching_2D}a shows the controlled invariant sets associated with the larger and smaller TEBs, and the over-approximated set associate with the SSB. The same information computed analytically is shown in Fig.~\ref{fig:TEB_Switching_2D}b, where the minimal SSB may be computed exactly.

\subsection{Online Meta-Planning}
\label{subsec:metaplanning_online}

At runtime, the meta-planner is in charge of constructing and maintaining a tree, $\mc{T}$, of waypoints connected via trajectories generated by the set of available planners.
It is also responsible for re-planning whenever new information about the environment becomes available, i.e. when obstacles are detected.

\begin{figure}[t]
\centering
\includegraphics[width=\columnwidth]{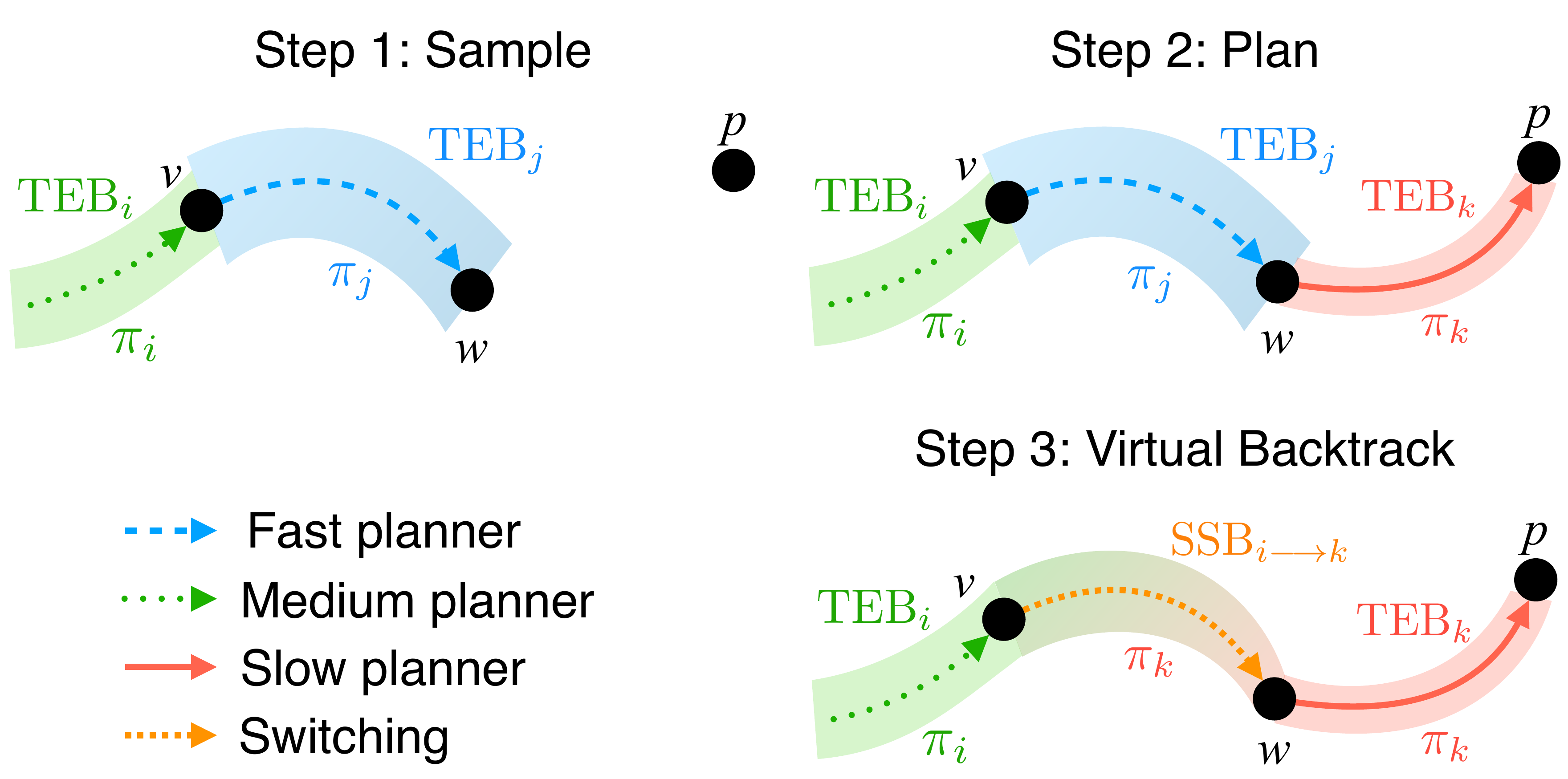}
\caption{Illustration of the online meta-planning algorithm.}
\label{fig:switching_logic}
\end{figure}

The precomputed safety sets allow the meta-planner to reason quickly about dynamic tracking feasibility as it builds~$\mc{T}$.
Using the precomputed TEBs, the meta-planner can determine which planners are safe to use in different regions of the environment.
In addition, the SSBs allow the meta-planner to determine the validity of planner-to-planner transitions. The meta-planner's logic is detailed below and illustrated in Fig. \ref{fig:switching_logic}.

\textbf{Step 0: Root.} The root node of $\mathcal{T}$ is initially set at
the starting position of the tracking system. 
Since the system has an initial tracking error equal to zero, it is by definition inside of all the available TEBs%
%and therefore the meta-planner can choose any planner to start with
. Later, if an obstacle has just been detected mid-trajectory, the new root node will be placed at the predicted position of the \emph{planning} system after some allowed computation time (typically $<$~$1$ s) and the \emph{tracking} system will only be guaranteed to be inside the TEB associated to the current edge of $\mathcal{T}$.

\textbf{Step 1: Sample.} The meta-planner constructs its tree $\mathcal{T}$ by sequentially sampling points uniformly at random from the environment and attempting to connect them to the nearest existing waypoint in the tree. %(the sampling can implement general efficient state-of-the-art algorithms \cite{gammell2014informed, gammell2015batch}).

\textbf{Step 2: Plan.} By default, the meta-planner always tries to connect waypoints using the fastest planner $\pi_1$, which is also associated to the largest TEB.
If $\pi_1$ does not find a collision-free trajectory, the meta-planner then attempts to use the second-fastest planner $\pi_2$, which has a smaller TEB.
The meta-planner continues trying available planners in order of decreasing TEB size until one succeeds or all have failed (in which case it abandons this candidate waypoint and samples a new one).

\textbf{Step 3: Virtual Backtrack.} When a planner $\pi_k$ succeeds in reaching a new point $p$
%\in\mathcal{E}$ 
from the nearest waypoint $w\in\mathcal{T}$, the meta-planner checks what planner was previously used to reach waypoint $w$ from its parent $v\in\mathcal{T}$.
If this preceding planner $\pi_j$ had a larger TEB than the new planner (that is, if $j<k$), then $p$ cannot be immediately added to $\mathcal{T}$.
Instead, the meta-planner first needs to ensure that the tracking system will be able to safely transition into $\text{TEB}_k$
\emph{before} reaching $w$, so that it can then track $\pi_k$'s plan from $w$ to $p$ while remaining inside its TEB%
%(which is what $\pi_k$ was assuming when trying to reach $p$
.
The meta-planner does this is by checking what planner $\pi_i$ was used to reach $w$'s parent $v$, and if $i < k$, using the safe switching bound $\text{SSB}_{i\to k}$
%from $\text{TEB}_i$ to $\text{TEB}_j$
to collision-check the already-computed path $v \to w$.
If $i \ge k$, there is no need to use a SSB and the path $v\to w$ is guaranteed to be safe under $\text{TEB}_k$, since it was already deemed safe under the larger $\text{TEB}_j$ by $\pi_j$.

If the check is successful, this means that, instead of getting from $v$ to $w$ tracking the faster planner $\pi_j$, the system can follow an alternative trajectory, skipping $\pi_j$ altogether and transitioning from the speed of $\pi_i$ to the speed of $\pi_k$.
This path is added to $\mathcal{T}$ as an \emph{alternative} to the original $v\to w$ path: the more-slowly-reached $w$ is a new node in $\mathcal{T}$, and $p$ is added to $\mathcal{T}$ as a child of this new node.

If the check is unsuccessful, the meta-planner does not add $p$ to the tree.
%At this point, different logics are possible. 
Two different options for handling this possibility are as follows: %briefly here noting their practical implications:
\begin{enumerate}[a)]
\item \textbf{Discard:} $p$ is discarded and the meta-planner moves on to sample a new candidate point.
\item \textbf{Recursive Virtual Backtrack:} the meta-planner marks $v$ as a waypoint that needs to be reached from its parent using a \emph{slower} planner than the original $\pi_i$, so that safe transition into $\text{TEB}_k$ will be possible. This will always be the case if $v$ is reached using $\pi_j$, since $v \to w$ is safe under $\text{TEB}_j\subset\text{TEB}_i$. Step 3 can then be repeated on $v$, and recursively applied (at worst) until the root of $\mathcal{T}$.
\end{enumerate}

One alternative option for handling planner-switching failures is to prevent them altogether by always using SSBs instead of TEBs for the planning in Step~2. In particular, replacing $\text{TEB}_i$ with $\text{SSB}_{i\to N}$ will ensure that planners will only attempt to add a candidate point $p$ to the tree if it would not only be possible to reach $p$ under this planner but also, if later deemed necessary, to do so while transitioning to the smallest TEB (so that subsequent nodes can be connected to it by any planner without the need for the backtracking verification in Step~3). The additional conservativeness introduced by this substitution depends on the relative tracker-planner dynamics, namely on how much larger $\text{SSB}_{i\to N}$ is than $\text{TEB}_i$.

\begin{remark}
In the case of a point-mass tracking model following a kinematic planner, we have $\text{SSB}_{i\to j} = \text{TEB}_i$, $\forall j>i$, and therefore this substitution does not need to be done explicitly nor does it introduce any additional conservativeness. The backtracking check in Step~3 is \emph{always guaranteed to succeed}.
\end{remark}

\begin{proposition}
Any plan generated by the meta-planner algorithm can be safely followed by the tracking system.
\begin{proof}
The proof is by construction of the meta-planner, based on FaSTrack guarantees; we provide an outline here. A point is only added to the meta-planning tree if there exists a sequence of planned trajectories that reach the point such that (a) each planned trajectory can be tracked by the system with an error bounded by the associated TEB, and is clear of known obstacles by at least TEB, (b) each transition between planners can be followed by the system with an error bounded by the corresponding SSB, and is clear of known obstacles by at least SSB, and (c) if new obstacles are detected, re-planning succeeds (at worst, a geometric planner can always reverse or stop) in time for the system to switch to the new plan before colliding.
\end{proof}
\end{proposition}
%To switch from the larger tracking error bound to the smaller one, we first must ensure that the switching safety bounds are collision-free. \SHnote{insert backtracking notes}. If computational resources are more important than conservativeness then an alternative algorithm simply expands the larger tracking error bound by its corresponding switching safety bound. This ensures the we can always switch between planners at any point in the online computation, and thus requires no backtracking.

%The meta-planning logic is outlined in Algorithm \ref{alg:backtrack}.
%\comment{Calling pseudo-code from here. Currently way too big (was formatted for single column).}
%\input{pseudocode} % Algorithm pseudocode

% ====================== RESULTS =====================
\section{Results}
\label{sec:results}

\begin{figure*}[h!]
\centering
\begin{subfigure}[t]{0.49\textwidth}
    \centering
    \includegraphics[width=\textwidth]{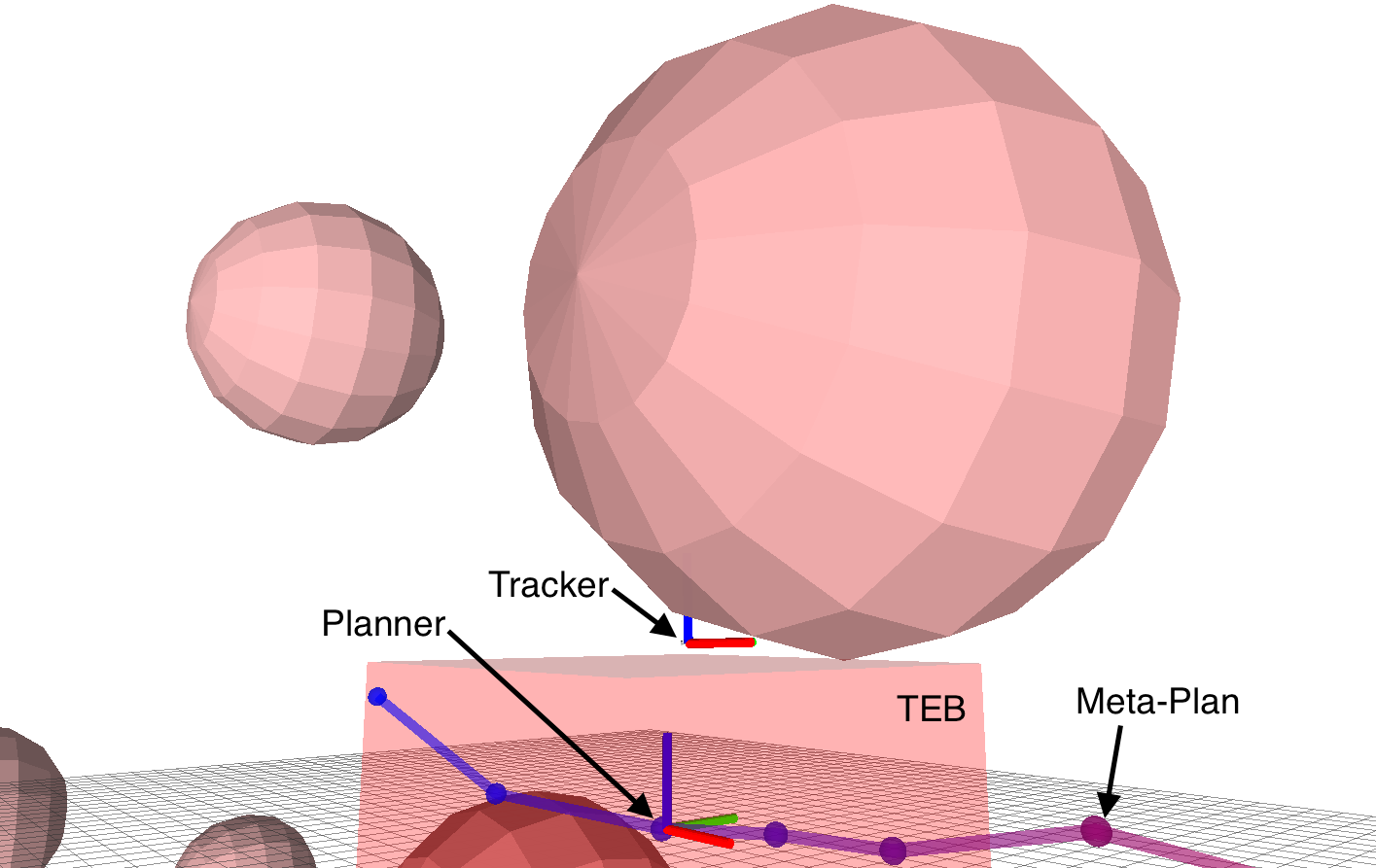}
    \caption{LQR controller.}
    \label{subfig:sim_lqr}
\end{subfigure}
~ 
\begin{subfigure}[t]{0.49\textwidth}
    \centering
    \includegraphics[width=\textwidth]{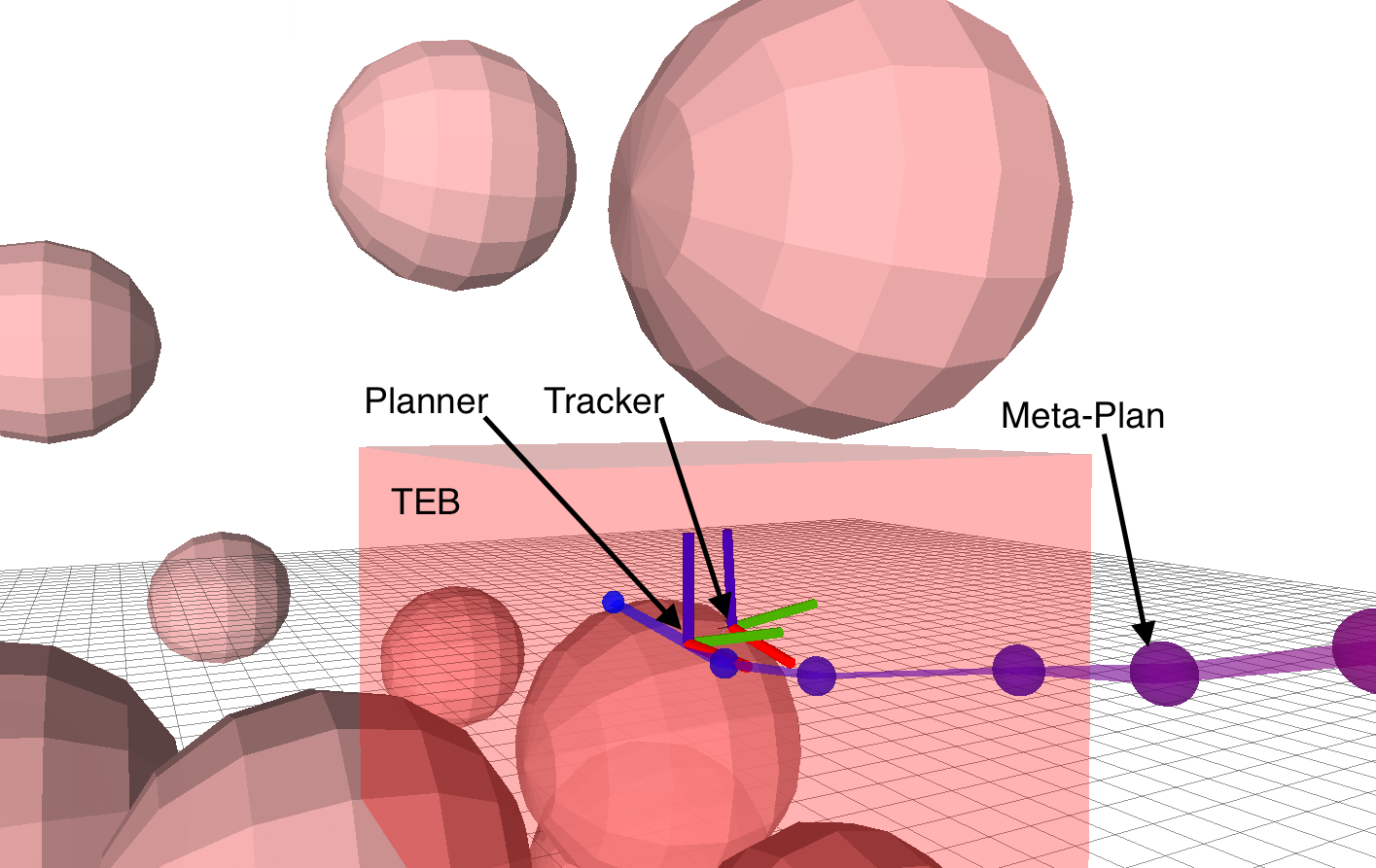}
    \caption{Safety controller.}
    \label{subfig:sim_opt}
\end{subfigure}
\caption{Simulated autonomous flight in a cluttered environment. Notice that when using LQR control the quadrotor leaves the TEB, but under optimal safety control it remains within the TEB. This is particularly important in the vicinity of obstacles.}
\label{fig:sim}
\end{figure*}

We demonstrate our algorithm on a 6D near-hover quadrotor model tracking a suite of 3D geometric planners running BIT* \cite{gammell2015batch} in the cluttered environment depicted in Fig.~\ref{fig:sim}  with different maximum speeds in each dimension. The tracking\footnote{Note that we have assumed a zero yaw angle for the quadrotor. This is enforced in practice by using a strict feedback controller on yaw rate to regulate yaw to zero.} and planning models (for the $i^{\text{th}}$ planner $\pi_i$) are given below in Eq. \ref{eq:Quad6D_dyn} (tracker at left, planner at right):

\noindent\begin{minipage}{.54\linewidth}
\begin{equation}
\small
\label{eq:Quad6D_dyn}
\begin{aligned}
\begin{array}{c}
\left[
\begin{array}{c}
\dot{\tstate}_{x}\\
\dot{\tstate}_{y}\\
\dot{\tstate}_{z}\\
\dot{\tstate}_{vx}\\
\dot{\tstate}_{vy}\\
\dot{\tstate}_{vz}\\
\end{array}
\right]
=
\left[
\begin{array}{c}
\tstate_{vx} - d_{vx}\\
\tstate_{vy} - d_{vy}\\
\tstate_{vz} - d_{vz}\\
g \tan\theta -d_{ax}\\
-g \tan\phi -d_{ay}\\
T - g - d_{az}
\end{array}
\right],
\end{array}
\end{aligned}\nonumber
\end{equation}
\end{minipage}\hspace{-0.8em}
\begin{minipage}{.45\linewidth}
\begin{equation}
\small
\label{eq:Quad3D_dyn}
\begin{aligned}
\begin{array}{c}
\left[
\begin{array}{c}
\dot{\pstate}_{x}\\
\dot{\pstate}_{y}\\
\dot{\pstate}_{z}\\
\end{array}
\right] =
\left[
\begin{array}{c}
b_x^{(i)}\\
b_y^{(i)}\\
b_z^{(i)} \\
\end{array}
\right]
\end{array}\\
\end{aligned}
\end{equation}
\end{minipage}\\

Here $\tctrl = [\theta, \phi, T]^T$ and correspond to roll, pitch, and thrust.  In all experiments, we set $-0.15 \mathrm{\ rad} \leq [\theta,\phi] \leq 0.15 \mathrm{\ rad
}$ and $7.81\ m/s^2 \leq T \leq 11.81\ m/s^2$. Planner $\pi_i$'s controls are $\pctrl = [b_x^{(i)}, b_y^{(i)}, b_z^{(i)}]$, each representing a fixed maximum speed in the given dimension. Due to the form of \eqref{eq:Quad6D_dyn}, the optimal safety controller will be bang-bang. However, it is only critical to apply the safety control at the \textit{boundary} of the TEB. A smooth linear controller may be used in the interior,
following a least-restrictive supervisory control law.
The relative dynamics between the tracking and planning models are:
\begin{equation}
\small
\label{eq:Quad6D_3D_dyn}
\begin{aligned}
\begin{array}{c}
\left[
\begin{array}{c}
\dot{\rstate}_{x}\\
\dot{\rstate}_{y}\\
\dot{\rstate}_{z}\\
\dot{\rstate}_{vx}\\
\dot{\rstate}_{vy}\\
\dot{\rstate}_{vz}\\
\end{array}
\right]
=
\left[
\begin{array}{c}
\tstate_{vx} - d_{vx} - b_x^{(i)}\\
\tstate_{vy} - d_{vy} - b_y^{(i)}\\
\tstate_{vz} - d_{vz} - b_z^{(i)}\\
g \tan\theta -d_{ax}\\
-g \tan\phi -d_{ay}\\
T - g - d_{az}
\end{array}
\right]
\end{array}
\end{aligned}
\end{equation}

%There is assumed to be an external acceleration disturbance in each dimension with bounds. %$-0.1 \hspace{2mm}\mathrm{ m/s^2} \leq [\dstb_{ax}, \dstb_{ay}, \dstb_{az}] \leq 0.1 \hspace{2mm}\mathrm{ m/s^2}$. There is also a velocity disturbance, which we assume is proportional to the planner velocity: $-\frac{1}{2}\pctrl \hspace{2mm} \mathrm{ m/s} \leq [\dstb_{vx}, \dstb_{vy}, \dstb_{vz}] \leq \frac{1}{2}\pctrl \hspace{2mm} \mathrm{ m/s}$. We introduce this scaled disturbance term \Snote{because of timing delays in the real system that will cause error proportional to the planning velocity}. 

\begin{figure}[h]
\centering
\includegraphics[width=0.45\textwidth]{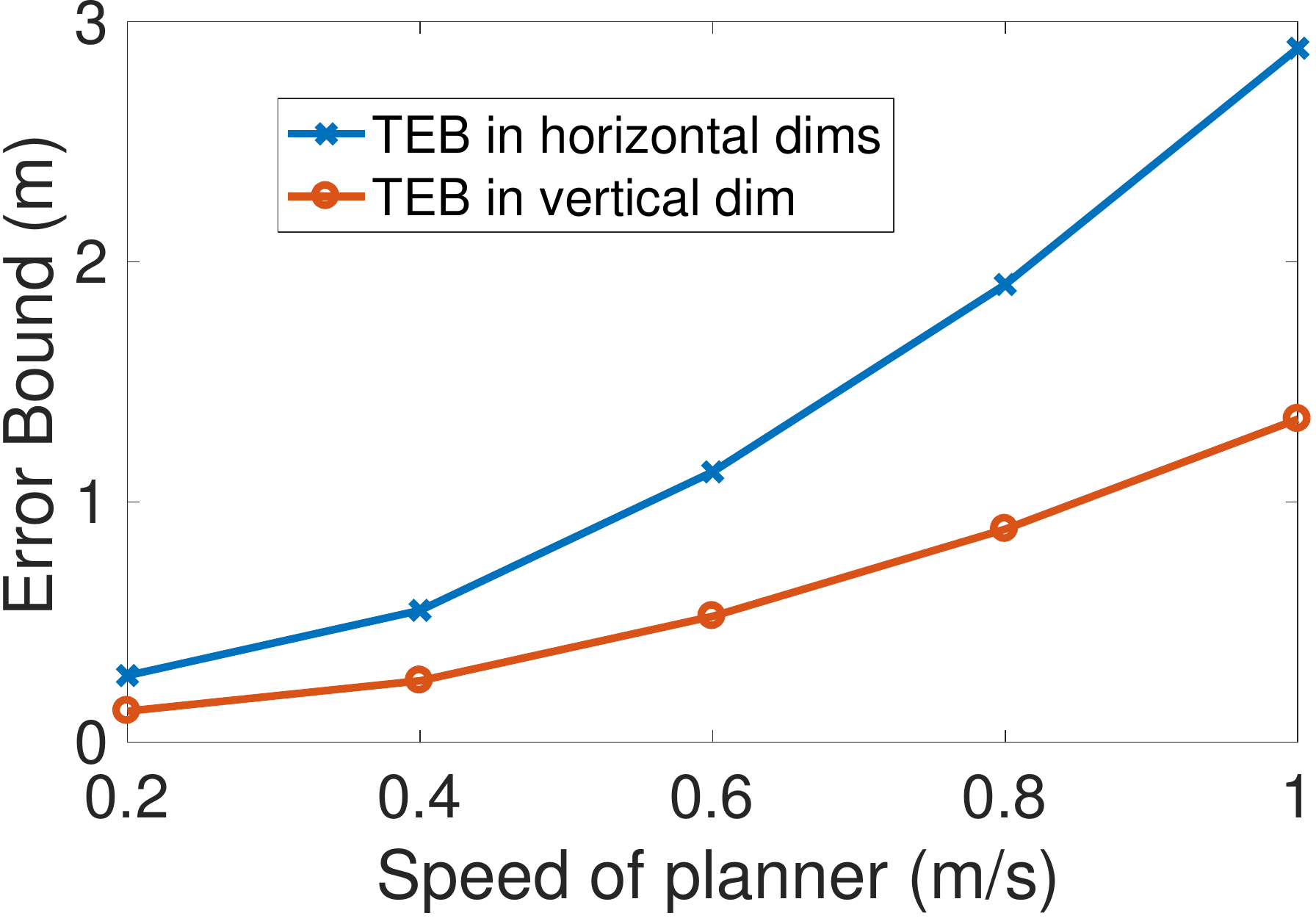}
	\caption{TEB vs. planner speed in each subsystem.}
	\label{fig:TEB_vs_Planner_Speed}
\end{figure}

Equation \eqref{eq:Quad6D_3D_dyn} can be split into three 2D subsystems with states $(x, v_x), (y, v_y),$ and $(z, v_z)$ that are of the same form as the double-integrator example from Section \ref{subsec:metaplanning_offline}. Note that the dynamics of the $(x,v_x)$ and $(y, v_y)$ subsystems are identical, and thus can be solved once and applied to each subsystem. By using decomposable HJ reachability \cite{Chen2016DecouplingJournal} we compute the $(x,v_x)$ set in 2~min 15~s and the $(z, v_z)$ set in 2~min, for a total of a 4~min 15~s precomputation time.
%\SHnote{add computed offline computation times}. Since the double-integrator affords an analytic solution for all quantities of interest (TEB, SSB, optimal controllers, etc.) we skip the offline computation step and simply compute these quantities in closed form online. 
Fig.~\ref{fig:TEB_vs_Planner_Speed} shows the growth of $\TEB_\pstate$ in each subsystem's position state as the planner speed in that dimension increases. Moreover, as explained in Section \ref{subsec:metaplanning_offline}, the TEB for $\pi_i$ is identical to the SSB for switching from $\pi_i \To \pi_j, j > i$.

\subsection{Simulation}
\label{subsec:simulation_example}

We implemented the meta-planning online algorithm within the robot operating system (ROS) \cite{ros} framework. We used the BIT* \cite{gammell2015batch} geometric planner from the Open Motion Planning Library (OMPL) \cite{sucan2012the-open-motion-planning-library}. Code is written in C++ and is available as an open source ROS package.\footnote{\href{https://github.com/HJReachability/meta\_fastrack}{\tt{https://github.com/HJReachability/meta\_fastrack}}} Meta-planning typically runs in well under one second in a moderately cluttered environment.

Fig.~\ref{fig:sim} shows a snapshot of a simulated autonomous quadrotor flight in an artificial environment with spherical obstacles using trajectories generated by our algorithm. Initially, the obstacle locations and sizes are unknown to the quadrotor, but as soon as they come within the sensing radius (the size of which must adhere to the constraint discussed in Section \ref{sec:background}) they are added to the meta-planner's internal environment model and used during re-planning. 

In Fig.~\ref{subfig:sim_lqr} we show what happens when the tracking controller is a standard LQR controller, while in Fig. \ref{subfig:sim_opt} everything remains the same except that we apply the optimal controllers derived from the offline analysis in Section \ref{subsec:metaplanning_offline}. Note that the LQR controller makes no guarantee about staying within the TEB, and hence it is unable to remain inside the TEB in the vicinity of the obstacle. The optimal controller, conversely, is guaranteed to remain in the TEB.

%Talk about how we simulated this in ROS, how the obstacles were represented, how sensing was done, screen shots of simulation (would be great to show one where we need to switch from the a bigger to a smaller bound).  Link to video of simulation.

\subsection{Hardware Demonstration}
\label{subsec:experiment_example}

\begin{figure}[ht]
\centering
\includegraphics[width=\columnwidth]{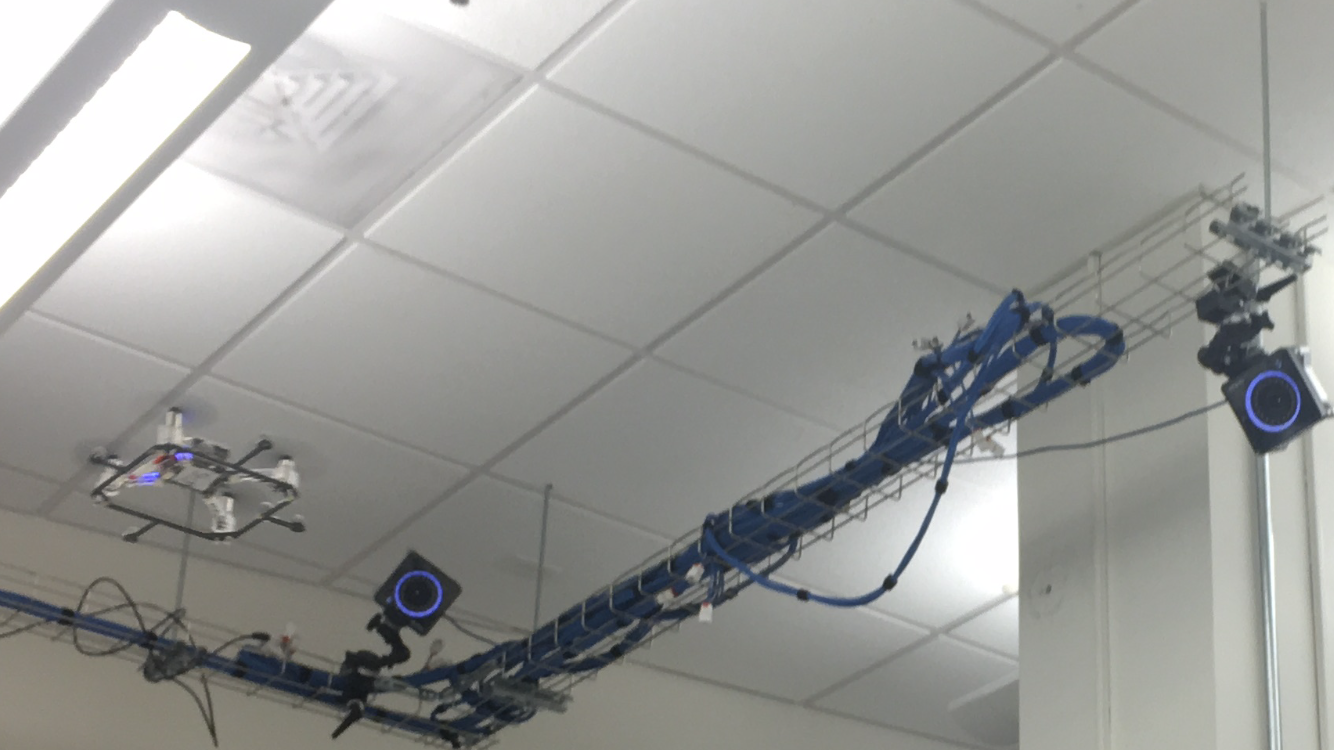}
\caption{A Crazyflie 2.0 flying during our hardware demonstration. Two OptiTrack cameras are visible in the background.}
\label{fig:crazyflie}
\end{figure}

We replicated the simulation on a hardware testbed using the Crazyflie 2.0 open source quadrotor platform, shown in Fig. \ref{fig:crazyflie}. We obtained position and orientation measurements at $\sim$~$235$ Hz from an OptiTrack infrared motion capture system. Given state estimates, we send control signals over a radio to the quadrotor at 100 Hz. As shown in our accompanying video,\footnote{\href{https://youtu.be/lPdXtR8Ar-E}{\tt{https://youtu.be/lPdXtR8Ar-E}}} the quadrotor successfully avoids the obstacles while remaining inside the TEB for each planner the meta-plan.

\begin{figure}[ht]
\centering
\includegraphics[width=\columnwidth]{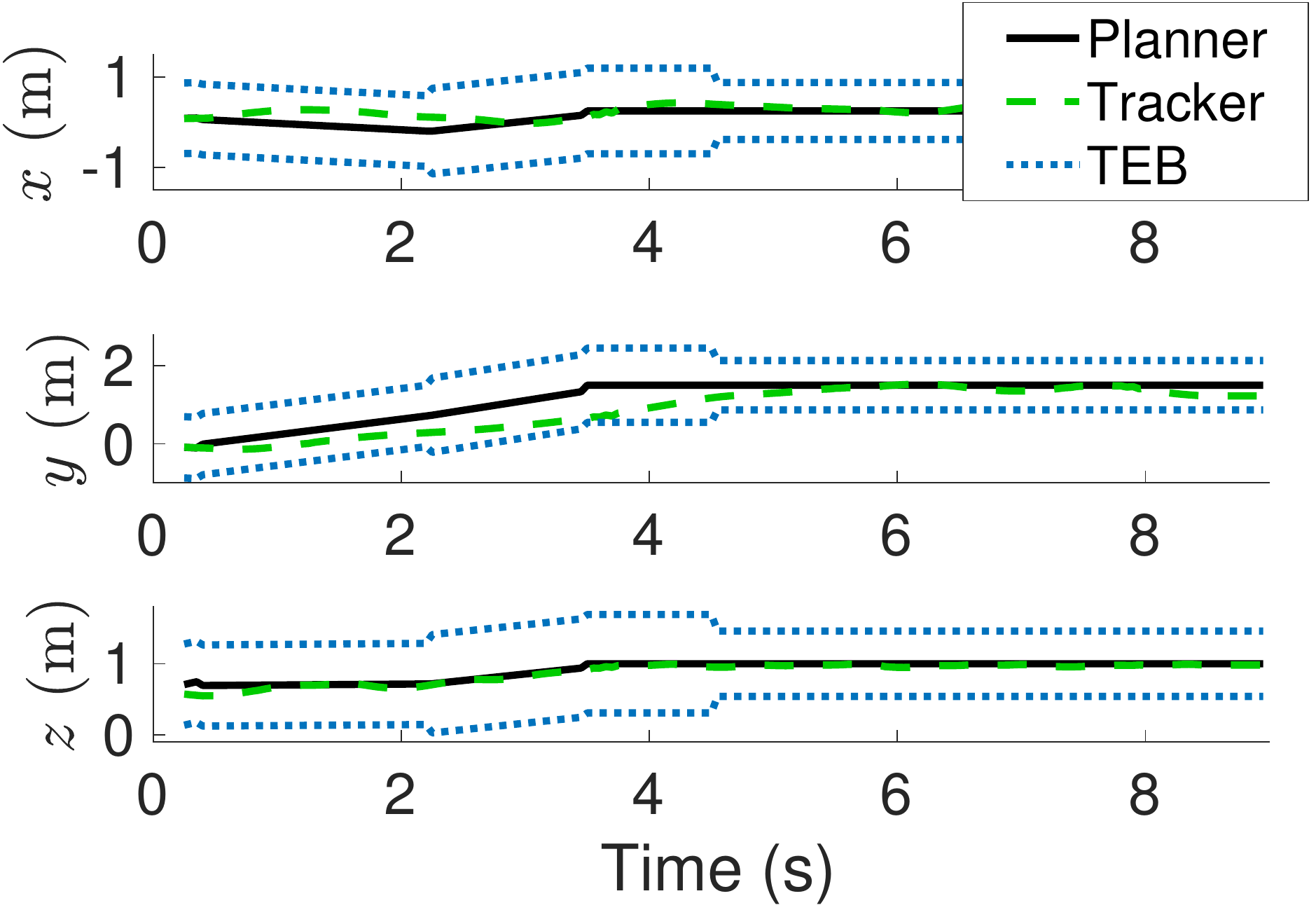}
\caption{Position vs. time during a hardware demonstration.}
\label{fig:hw_data}
\end{figure}

Fig.~\ref{fig:hw_data} shows the quadrotor's position over time recorded during a hardware demonstration. Note that the quadrotor stays well within the TEB throughout the flight even when the TEB changes size during planner switches.

%Talk about how we did this on the Crazyflie 2.0.  Discuss optitrack, obstacles and their representation.  picture of crazyflie moving through obstacles, link to video. 

% ===================== CONCLUSIONS ===================
\section{Conclusions}
\label{sec:conclusions}

We have proposed a novel meta-planning algorithm for using FaSTrack with multiple planners. The algorithm adaptively selects the fastest-moving planner that finds locally collision-free paths, and guarantees safe online transitions between these planners. The resulting meta-plans use more aggressive, faster-moving planners in open areas and more cautious, slower-moving planners near obstacles. We demonstrate meta-planning in simulation and in a hardware demonstration, using a quadrotor to track piecewise-linear trajectories at different top speeds.

The theory we develop here is general and can be applied to a wide variety of systems, including manipulators and other mobile robots. However, computing the TEB and SSB using HJ reachability can be challenging for these high-dimensional coupled systems. Ongoing work seeks to alleviate this challenge using other methods of computation such as sum of squares programming and neural network function approximators. Other promising directions include incorporating time-varying obstacle avoidance, further integration with OMPL and other planning libraries, providing adaptable error bounds based on external disturbances, and updating the tracking error bound online based on learned information about the tracking model.
%We are excited to explore several future directions of meta-planning with FaSTrack in the future, including integration with the Open Motion Planning Library (OMPL) and other planning libraries to allow for a greater variety of planning method. 

% BIBLIOGRAPHY =============================================%
% \addtolength{\textheight}{-12cm} % put on page before last.
% \section*{APPENDIX}
% \section*{ACKNOWLEDGEMENT}
\balance
\printbibliography
\end{document}